# Computational study of geometric effects of bottom wall microgrooves on cell docking inside microfluidic devices

**Sina Ahandoust[1]**
[1]*Department of Biomedical Engineering,*
*Amirkabir University of Technology*
*424 Hafez Ave, Tehran, Iran.*
*sinaahandoust2@gmail.com*

**Maryam Saadatmand[2]**
[2]*Department of Chemical and petroleum Engineering,*
*Sharif University of Technology*
*Azadi Ave, Tehran, Iran.*
*m.saadatmand@sharif.edu*
[2]*Corresponding author*

Cells docking inside microfluidic devices is effective in studying cell biology, cell-based biosensing, as well as drug screening. Furthermore, single cell and regularly cells docking inside the microstructure of microfluidic systems are advantageous in different analyses of single cells exposed to equal drug concentration and mechanical stimulus. In this study, we investigated bottom wall microgrooves with semicircular and rectangular geometries with different sizes which are suitable for single cell docking along the length of the microgroove in x direction and numerous cells docking regularly in one line inside the microgroove in a 3D microchannel. We used computational fluid dynamics to analyze the fluid recirculation area inside different microgrooves. The height of recirculation area in the bottom of microgroove can affect the cell's attachment, and also materials delivery to attached cells, so the height of recirculation area has to be optimum amount. In addition, we analyzed the fluid drag force on cell movement toward the microgroove. This parameter was proportional to the fluid velocities in x and y directions changing in different microgrooves geometries. In different microgrooves' geometries the fluid velocity in y direction does not change. If the fluid velocity in x direction decreases inside the microgroove, the cell movement time inside the microgroove will increase, and also the drag force in y direction can push the cells toward the bottom due to the lower drag force in x direction. The percentages of negative shear stress and average shear stress on the adhered cell surface were also calculated. The lower average shear stress, and negative shear stress around 50% on the cell surface are against cell detachment from the substrate. The results indicated that at the constant fluid inlet velocity and microchannel height, microgroove geometry and ratio of cell size to the microgroove size play pivotal roles in the cell initial adhesion to the substrate as well as the cell detachment.

# 1 Introduction

Docking and capturing of cells are critical in various biomedical applications including diagnostics, cell biology, and therapeutics [1]. Nowadays, single cell analysis has taken precedence over studying bulk population of cells due to discovering mechanisms related to genomics, proteomics, and metabolomics. Specifically, single cell analysis has advantages such as analyzing the cell antigen and chromosome, exposure to PHAs and analyzing RNA and DNA alteration, analyzing cytotoxicity of the cell and different cell fates [2]. In some biomedical applications, capturing irregular accumulation of cells cannot offer suitable results, so the experiment needs orderly and organized cell docking in order to achieve suitable results. Some of these biomedical applications involve studying cell mechano-transduction including different cells sense and response to mechanical stimuli [3, 4], embryonic stem cell undifferentiated growth, and to form homogenous embryonic stem cell aggregates to enhance their differentiation for therapeutic purposes [5, 6]. Another usage includes studying cellular interactions and adhesion [7], and providing multiphenotype cell arrays for drug discovery experiments[2, 8]. Some conventional techniques used for cell sorting and capturing are Fluorescence-activated cell sorting (FACS) and Magnetic-activated cell sorting (MACS), which utilize complementary fluorophore-conjugated antibodies to label cells of interests and magnetic beads to bind specific proteins on cells of interest, respectively. These two conventional techniques with labels may affect cell fate and function, which both consume costly reagents [9].

Nowadays, microfluidic devices have different biological and biochemical applications. These microdevices can manipulate fluid flows, enable high-throughput experimentation while minimizing costly reagent consumption and reducing sample processing time[10, 11]. For capturing cells within microfluidic channels, a number of methods have been widely used such as encapsulation within photocrosslinkable polymers and use of magnetophoresis, dielectrophoresis, and acoustophoresis, which may adversely affect cell biology studies [12-16]. However, target cell sorting and docking due to fluid hydrodynamic force inside the microchannel do not affect cell biology studies and cell phenotype [17]. Since hydrodynamic forces on the target cells, for example leukocytes, inside the microchannel are similar to the physiological condition inside the body (in-vivo), so the cells experience almost the same conditions in-vitro [18].

Some previous studies on cell docking inside microfluidic channels have included substrates with microgrooves or microwells in order to improve cell docking [8, 19-23]. These studies have been experimentally conducted through cell seeding inside microfluidic systems. Due to the high cost of experiments, they have only studied rectangular microgrooves for cell positioning. Recent studies of the role of microgrooves in cell docking have been accompanied by simulation of fluid shear stress and recirculation area at the bottom of microgrooves before cell seeding [19-21]. In these studies, however, the interaction between fluid and cells has not been simulated. In addition, they have studied rectangular microgrooves with large dimensions for capturing many cells inside microgrooves; therefore, irregular accumulation of cells inside a microgroove cannot provide the condition for aforementioned biomedical applications requiring organized cell docking.

In this work, we studied cell docking in a flat microchannel and microchannels with semicircular and rectangular microgrooves with different dimensions. The assumed microgrooves dimensions were suitable for single cell docking in a x direction along the length of the microgroove and for cells positioning in one line in a 3D microchannel. We used computational simulation to obtain the fluid velocity inside different microgrooves and shear stress on the cells attached inside the microgrooves. Computational fluid dynamics (CFD) revealed the effect of two kinds of microgrooves with different sizes on the fluid velocity and flow direction, which are related to the drag force on the moving cell. In addition, fluid-structure interaction (FSI) modeling determined the fluid shear stress distribution on the adhered cell to predict the cell detachment. In this work, numerical simulation of microfluidic devices has the privilege of assessing the microdevice geometrical features and analyzing mechanical stress on the cells before the fabrication; therefore, it can reduce the cost of fabricating different microdevices.

## 2 Models and Methods

### 2.1 *Microgrooves' Geometries*

In this work, a microchannel with flat bottom wall and two kinds of grooved bottom walls have been studied. The assumed microgroove sizes were appropriate for docking single cell in the 2D microchannel or numerous cells regularly in a line in the 3D microchannel. In other words, the assumed microgrooves' sizes were only for one cell docking in x direction along the microgroove's length as shown in Fig. 1. Thus, the docked cells in a microgroove could be exposed to same drug and material concentration. The target cells were spherical with a radius of 10 μm. In a microfluidic device with parabolic velocity profile, spherical cells such as leukocytes are driven closest to the wall where the flow velocity is minimum [24]. According to Table 1, the radius of the assumed semicircular microgrooves was between 12 μm and 20 μm. In the microchannel with rectangular microgrooves, the height of rectangles was between 12 μm and 20 μm, and for each height we assumed two lengths one of which was equal to the length or diameter of the corresponding semicircular microgroove, while the other was representative of the same areas of corresponding semicircular and rectangular microgrooves. Fig. 1 demonstrates flat microchannel and microchannels with two kinds of microgrooves (rectangular and semicircular) where the length, height, and width of the microchannel are along x, y, and z directions, respectively.

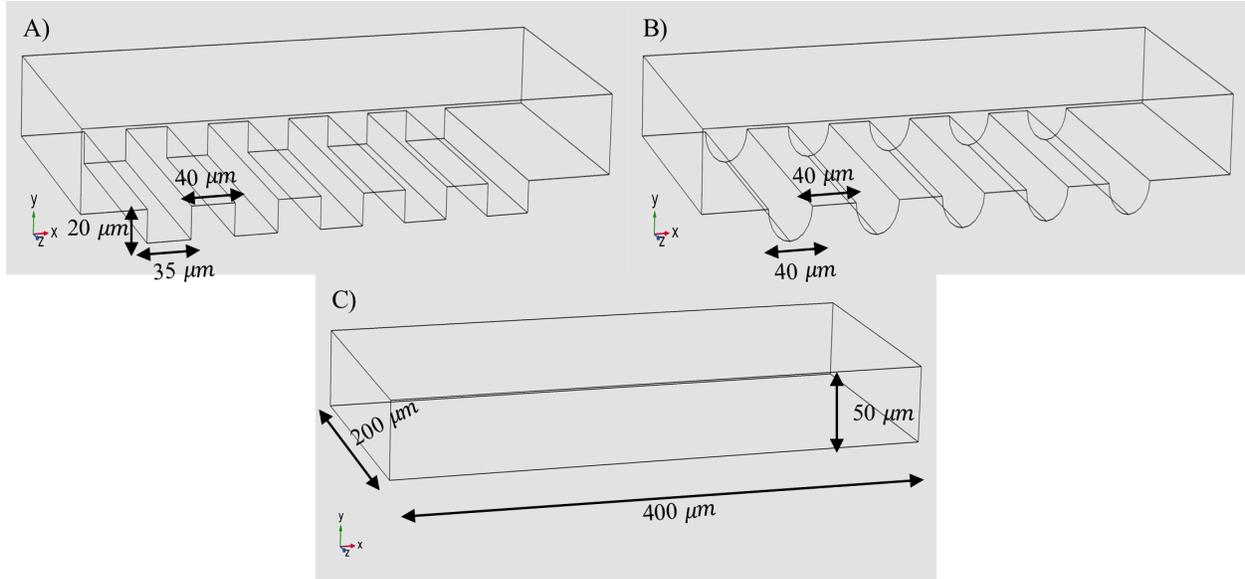

Fig. 1. 3D view of microchannels with same dimensions (400 μm×50 μm×200 μm) ; A)Rectangular microgrooves (20 μm ×35 μm), B) Semicircular microgrooves (r= 16 μm), and C) Flat microchannel without microgrooves

Table 1. Dimensions of assumed microgrooves at the bottom wall of microfluidic devices

| Semicircular microgroove radius (μm) | Rectangular microgroove dimensions (height×length (μm$^2$)) |
|---|---|
| r=12 | s=12×24 |
|  | s=12×20 |
| r=14 | s=14×28 |
|  | s=14×22 |
| r=16 | s=16×32 |
|  | s=16×26 |
| r=18 | s=18×35 |
|  | s=18×29 |
| r=20 | s=20×35 |

The height and length of the microchannel were assumed 50 μm and 400 μm, respectively, and the distance between two microgrooves was 40 μm. As presented in Fig. 2, 2D computational domain was discretized into an unstructured mesh of triangular elements. The maximum mesh length was set to 2 μm in the bulk and 0.8 μm at all corners and microgroove edges; it was confirmed that the results of simulation were grid-independent.

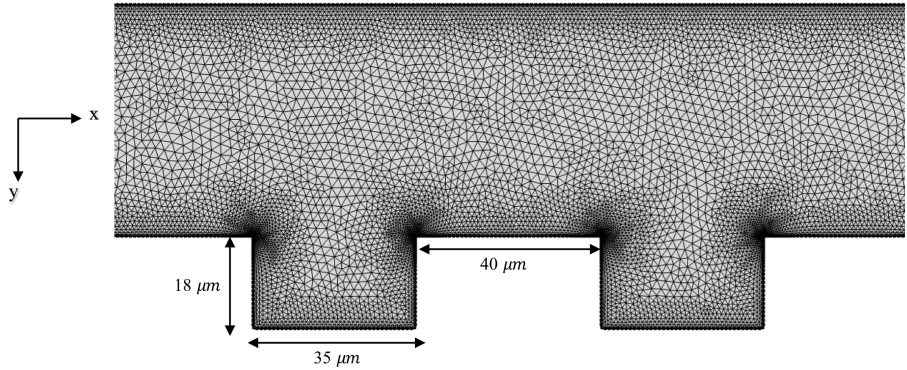

Fig. 2. Mesh structures with triangular elements for microchannel with rectangular microgrooves with height of 20μm and length of 35μm.

## 2.2 Numerical simulation of fluid flow

In this part, 2D computational fluid dynamics (CFD) was used to determine the fluid recirculation area and velocity direction inside the microgrooves by COMSOL Multiphysics 5.3. The laminar fluid flow was assumed Newtonian, incompressible, and homogenous. The steady state Navier-Stokes equations for incompressible fluid were defined as:

$$F + \nabla . [-PI + \mu(\nabla u + (\nabla u)^T)] = \rho((u).\nabla)u \quad (1)$$
$$-\nabla . u = 0 \quad (2)$$

Where, $I$ denotes the unit diagonal matrix, $F$ is the volume force affecting the fluid due to the gravity force, $u$ is velocity vector, and $P$ is fluid pressure. The fluid was assumed as water with viscosity of $\mu$=8.9×10$^{-4}$Pa.s and density of $\rho$=1000Kg/$m^3$.

The boundary conditions at the walls and at the bottom of the microgrooves were set as no-slip boundary condition. The specified velocity condition equal to 2×10$^{-4}$m/s was applied for the inflow boundary condition. Further, the specified pressure of 0 Pa was used for the outflow boundary condition. Eventually, the criterion for convergence (RMS residual) was considered to be equal to 10$^{-6}$.

## 2.3 Analyzing fluid drag force on the moving cell toward the microgroove

The moving cells experience different forces from the fluid including drag force, buoyance force, and weight [25].Weight and buoyancy force are independent of microgroove geometry and were assumed constant in different microchannels. Fig. 3 displays all the forces acting on the cell when the cell is about to enter the microgroove. In order to compare the moving cell toward different geometries of microgroove, the drag forces in x and y directions are important as other forces are constant in different geometries.

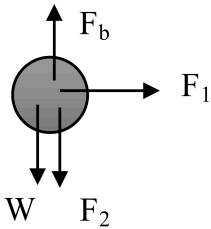

Fig. 3. The force diagram on the moving cell in the fluid [25]

In Fig. 3, $F_1$ and $F_2$ are drag forces in x and y directions, respectively. The drag force on the moving sphere with radius a in the laminar flow can be defined as [25]:

$$F_{drag} = \int_S n.\sigma \, dS = 6.\pi.\mu.a.V \quad (3)$$

In Equation (3), $\sigma$ is the fluid stress tensor which has contributions from both pressure and viscous stresses and **n** is the unit normal direction from sphere surface into the fluid domain. The third side of Equation (3) is known as the Stokes drag formula, and $\mu$, $a$, and $V$ denote fluid dynamic viscosity, radius of sphere, and speed of the sphere related to the fluid, respectively. So, the drag force is proportional to $V$. Therefore, we analyzed velocities in x and y directions and illustrated the different drag forces on the cell when the cell was entering different microgrooves.

### 2.4 *Analyzing the cell rolling on the microchannel substrate*

When the cell is initially adhering to the coated receptors of the surface, the adhesion force is important in stopping the cell rolling in contrast to the fluid shear stress. Former experimental studies have discovered the adhesion force of different types of cells such as different leukocytes ligands and special proteins (receptors) in-vitro by aspiration force of a micropipette [26, 27]. The normal stress distribution acting on the interface between cell ligands and surface receptors should be balanced by the resultant fluid shear force acting in the center of the cell for cell attachment. When the resultant shear force due to fluid shear stress on the cell surface is higher than the micropipette aspiration force causing cell detachment, the cell will be detached from the surface [27, 28]. So, we analyzed the absolute cell detachment from surface by comparing experimental aspiration force with the resultant shear force on the cell in different positions and conditions inside the microfluidic device.

### 2.5 *Numerical simulation of fluid and the adhered cell interaction*

Here, 2D finite element method by COMSOL Multiphysics 5.3 was used to study the fluid-structure interaction (FSI) including fluid flow shear stress effect on cells adhered in microchannels with different microgrooves. The steady state Navier-Stokes equations for incompressible fluid have been defined in Equations (1) and (2). The cells were modeled elastic due to the steady state flow, where the shear stress did not fluctuate over time. The cell attachment smoothing at the juncture with substrate was assumed with a fillet of 0.2 cell radius (2μm)[29]. The equation of viscous and pressure forces of fluid on the cell was defined as:

$$F_T = -n.(-PI + \mu(\nabla u + (\nabla u)^T) \quad (4)$$

Where, $n$, $F_T$, and $u$ represent the normal vector to the boundary, the sum of viscous and pressure forces, and fluid velocity field, respectively [30].
In Equation (5), $F$ denotes the deformation gradient tensor which presents the relationship of a material line $dX$ before deformation to the line $dx$ after deformation, $\nabla$ refers to the gradient operator with respect to $x$, and u is the displacement vector of the particle inside the body.

$$F = \left(\frac{\partial x}{\partial X}\right)^T \equiv (\nabla x)^T = (\nabla u + I)^T \quad (5)$$

Equations of stress on the cell can be defined as:

$$df = S.dA = F^{-1}.d\dot{f} \quad (6)$$
$$J = \det(F) \quad (7)$$
$$S = J.F^{-1}.\sigma.F^{-T} \quad (8)$$

Where, the second Piola-Kirchhoff stress tensor (*S*) is introduced as the stress tensor associated with the force df in the undeformed elemental area (*dA*). In addition, the force $d\dot{f}$ on the deformed elemental area (*da*) is related to the force *df* on the undeformed elemental area (*dA*). In Equation (8), the second Piola-Kirchhoff stress is related to Cauchy stress tensor σ [31].

The elasticity equations of the isotropic and symmetric cells can be defined as [31]:

$$\varepsilon_{ij} = \frac{1}{E}((1+\nu)\sigma_{ij} - \nu\delta_{ij}\sigma_{kk})) \quad (9)$$

Where, $\varepsilon_{ij}$, $\sigma_{ij}$, and $\sigma_{kk}$ are strain tensor, component of deviatoric stress tensor, and hydrostatic stress tensor, respectively [31]. In this study, Poisson ratio and Young modulus of the cell were assumed as $\nu = 0.3$ and $E = 300$ Pa [32].

## 3 Results and Discussion

### 3.1 Fluid flow recirculation area

Recirculation of fluid flow can have a profound effect on the cells attachment to the coated proteins on the microchannel surface. This assumption was also proved experimentally in the former studies inside the microwells in a microfluidic channel [6, 20]. We assessed the recirculation area parameters across different dimensions of two kinds of microgrooves through numerical simulation.

Fig. 4 reveals the fluid streamlines in different microgrooves. It is concluded that the flow recirculation area inside the microgroove is related to the height and length of the microgroove. In the rectangular microgrooves, fluid recirculation can occur everywhere at the bottom of the microgrooves (Fig. 4A(i)) or only on the corners of the microgrooves (Fig. 4A(ii)). Recirculation at the entire bottom of the microgroove can significantly affect cell attachment, while the recirculation on the corner does not improve the cell attachment to the surface. In semicircular microgrooves, the recirculation area is at the bottom of microgroove, but its height depends on the radius of the microgroove (Fig. 4B(i,ii)). If the recirculation area is deep and narrow inside the microgroove, the recirculation does not improve the cell attachment to the surface (Fig. 4B(ii)).

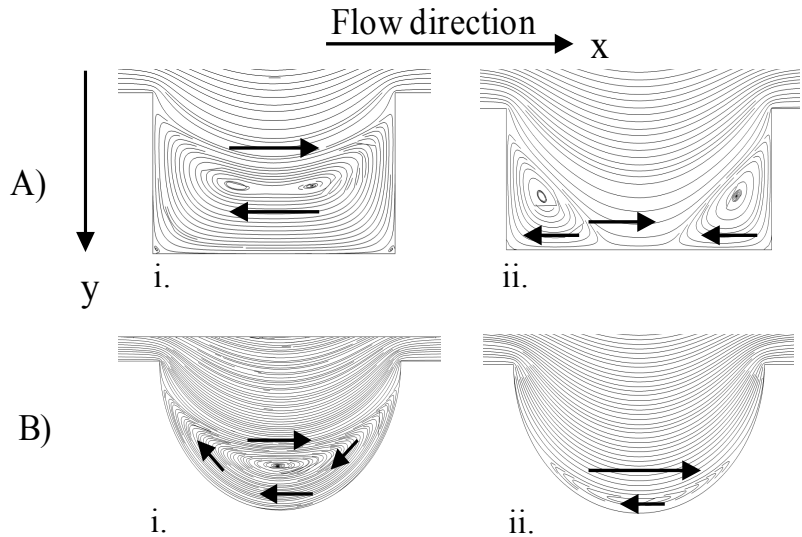

Fig. 4. Fluid streamlines and recirculation area inside the semicircular and rectangular microgrooves. The flow direction and assumed coordinates x and y are shown. In each microgroove, the flow direction is shown by arrows; A) Rectangular microgroove dimensions (height×length) i) 14×35 μm², ii) 16×26 μm²; B) Semicircular microgroove radius i) 18μm, ii) 24μm.

Fig. 5 demonstrates our defined parameters of recirculation area in semicircular and rectangular microgrooves. Parameter (a) is the maximum height of recirculation area near the edge of the microgroove. Parameter (b) shows the minimum height of the recirculation area in the center of the microgroove. Table 2 presents the recirculation parameters in the six microchannels with semicircular microgrooves which have different radii. At all of the assumed radii, we had recirculation area, and when the radius was lower than 12 μm, the radius of assumed cell was as same as microgroove's radius; therefore, the probability of cell docking inside the microgrooves was very low given the assumed cell radius (10 μm). The semicircular microgroove radius higher than 20 μm allowed for two cells docking in x direction, which also had a narrow recirculation area (as illustrated in Fig. 4B(ii)). Table 2 also represents the recirculation parameters in the 9 microchannels with rectangular microgrooves which have different dimensions. The height of the rectangles (from 12 μm to 20 μm) are equal to the radius of the semicircle microgrooves. Note that for each height of microgrooves, the length larger than the maximum assumed lengths did not have fluid flow recirculation at the entire bottom surface of the microgrooves. We also assessed the deeper rectangular microgrooves, but they can whelm the cell because of high recirculation area.

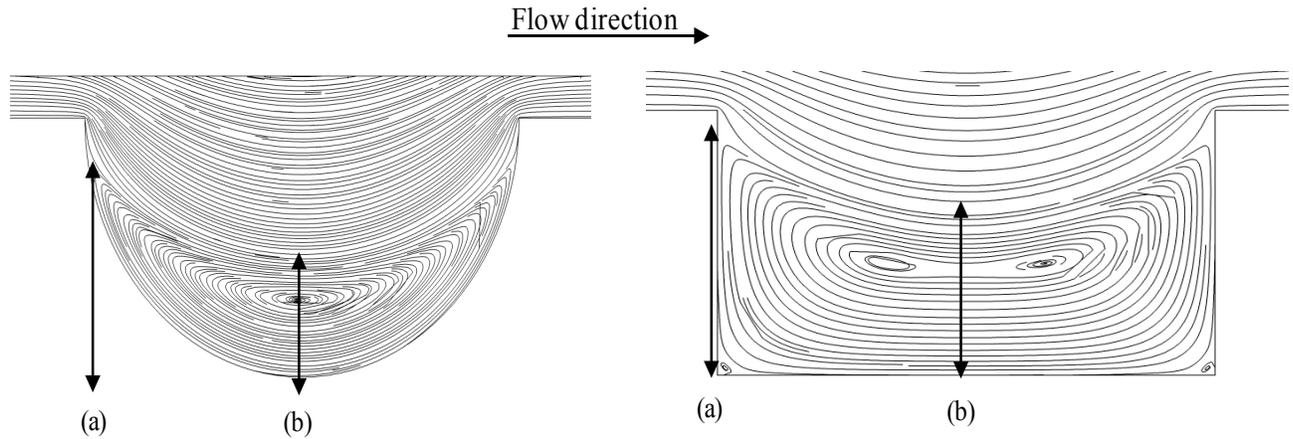

Fig. 5. Velocity streamline showing the fluid flow recirculation area inside the semicircular and rectangular microgrooves. (a) and (b) represent the recirculation area parameters.

Table 2. Maximum and minimum height of recirculation area inside semicircular microgrooves with different radiuses, and rectangular microgrooves with different dimensions

| Semicircular microgroove radius (μm) | a (μm) | b (μm) |
|---|---|---|
| 12 | 10.4 | 5.8 |
| 14 | 11.3 | 6.2 |
| 16 | 12.2 | 6.4 |
| 18 | 13.5 | 7.1 |
| 20 | 14.2 | 5.6 |

| Rectangular microgroove dimensions (height×length μm$^2$) | a (μm) | b (μm) |
|---|---|---|
| 12×20 | 11 | 8.5 |
| 12×24 | 11 | 6.8 |
| 14×22 | 13.8 | 7.9 |
| 14×28 | 12.5 | 7.6 |
| 16×26 | 14.6 | 11.7 |
| 16×32 | 14.6 | 11.7 |
| 18×29 | 16.4 | 12.7 |
| 18×35 | 16.4 | 12.7 |
| 20×35 | 17.7 | 12 |

Higher recirculation areas can enhance the possibility of cells detachment from fluid flow and cells attachment to the surface receptors in microgrooves. In other words, higher recirculation area inside the microgroove can cause a large separated region of fluid streamlines as well as cells separation from fluid flow. On the other hand, the height of recirculation area has an important role in delivering nutrients to the docked cells and cellular waste disposal, which are important for survival of cells. Higher recirculation area can whelm a larger volume of cells and reduce the delivery of nutrients to the cells as well as cellular waste disposal. So, the recirculation areas' heights should not be too high. The maximum height of the recirculation area is close to edges of the microgroove. When the maximum height of recirculation area (a) is high, A large portion of the cell can experience the fluid flow recirculation pushing the cell to the left side of microgroove, i.e. the negative velocity direction of the fluid flow in the recirculation area can push the cell to the left side of the microgroove. When the cell lies inside the microgroove, the entire surface of the cell should not be immersed in the recirculation flow for receiving nutrients and materials; so the maximum height of the recirculation area should be lower than the cell diameter. According to the cell radius (10 μm) and the attachment smoothing at the juncture with substrate by a fillet of radius 2 μm, the maximum recirculation height should be lower than 18 μm, so the maximum height of assumed rectangular microgroove is 20 μm. Based on comparing the two assumed rectangular microgrooves with the same heights and different lengths, the maximum and minimum heights of recirculation showed a minor difference. In the following, in rectangular microgrooves, the cell can move longer time in a recirculation area inside the microgroove with a higher length, so this can lead to higher possibility of cell docking in the microgroove with a large length. Therefore, for accelerating the computations, we assume only five rectangular dimensions with higher lengths from the nine assumed dimensions in Table 2.

### 3.2 *Fluid drag force on the moving cell toward the microgroove*

In order to compare the moving cell toward microgrooves of different geometries, the drag forces in x and y directions are important. According to Equation (3), by changing *V* in different microgrooves, the drag force on the cell will change. When the cell enters the microgroove, changing velocity and streamlines pattern can play important roles in changing drag forces. Fluid velocities in x and y directions have an effect on the cell moving to the bottom of the microgroove and its attachment. We assessed the velocity change in the streamline near the wall going into the microgroove in order to compare drag forces in different geometries. The origin of the assumed streamline is near the edge of the microgroove; the cell bottom is located in this streamline. Other parallel streamlines which cause the cell motion had variations similar to the assumed streamline change. As shown in Fig. 6, we assumed five points in the streamline entering the microgroove to assess the velocity change of fluid entering the different microgroove geometries in order to compare the fluid drag forces in different microgrooves. In all geometries, the assumed streamline entering inside each microgroove had 0.2 μm distance from the edge of microgroove, so all selected points had the same conditions across all geometries, and the first assumed point was 0.2 μm above the edge of microgroove as shown in Fig. 6 . The assumed points had also about 2 μm difference in y direction, and in rectangular microgrooves with heights 12 μm and 14 μm the assumed streamline did not go 8 μm in microgroove's depth in y direction as shown in Fig. 6; therefore, there were 4 assumed points for these geometries. The little velocity difference in y direction in all geometries can be negligible, so the fluid drag forces in y direction in different microgrooves are almost same. Fig. 6C represents the schematic view of the cell and fluid streamlines which are entering the assumed rectangular microgroove. Fig. 7A represents the column chart of the assumed streamline velocities in x direction ($V_x$) at five points in semicircular microgrooves. Fig. 7B displays the length of the last point in the assumed streamline in x direction representing the length of cell moving inside different microgrooves at the same height of assumed points. Fig. 8A shows the column chart of the assumed streamline velocities in x direction ($V_x$) at five points in rectangular microgrooves. However, in two smaller microgrooves, four points were assumed due to the lower height of the assumed streamline. Fig. 8B indicates the length of the last point in the assumed streamline in x direction except for two smaller rectangular microgrooves. This length shows the assumed streamline pattern, which is related to the cell movement inside the microgroove.

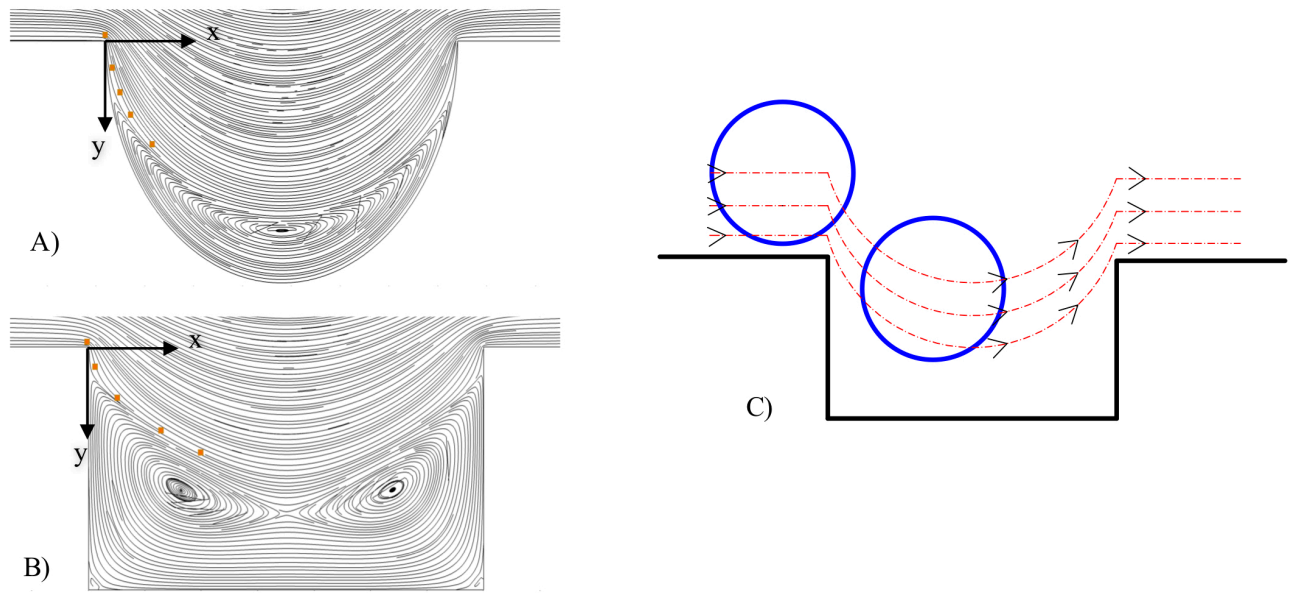

Fig. 6. A) Five assumed points in the streamline entering semicircular microgroove with radius 18μm. B) Five assumed points in the streamline entering the rectangular microgroove with height 18μm and length 35μm. Both sample microgrooves have the same height and area. C) Schematic view of the cell with fluid flow streamlines entering the rectangular microgroove

Lower velocity in x direction causes lower drag force on the cell in x direction. If the drag force in x direction decreases, the cell will move toward the bottom of the microgroove due to the drag force in y direction and cell weight. So, this can help in cell attachment to the substrate. In addition, the lower velocity in x direction causes increase cell moving time in x direction, so the cell can remain for a longer time inside the microgroove, thus increasing the probability of cell attachment.

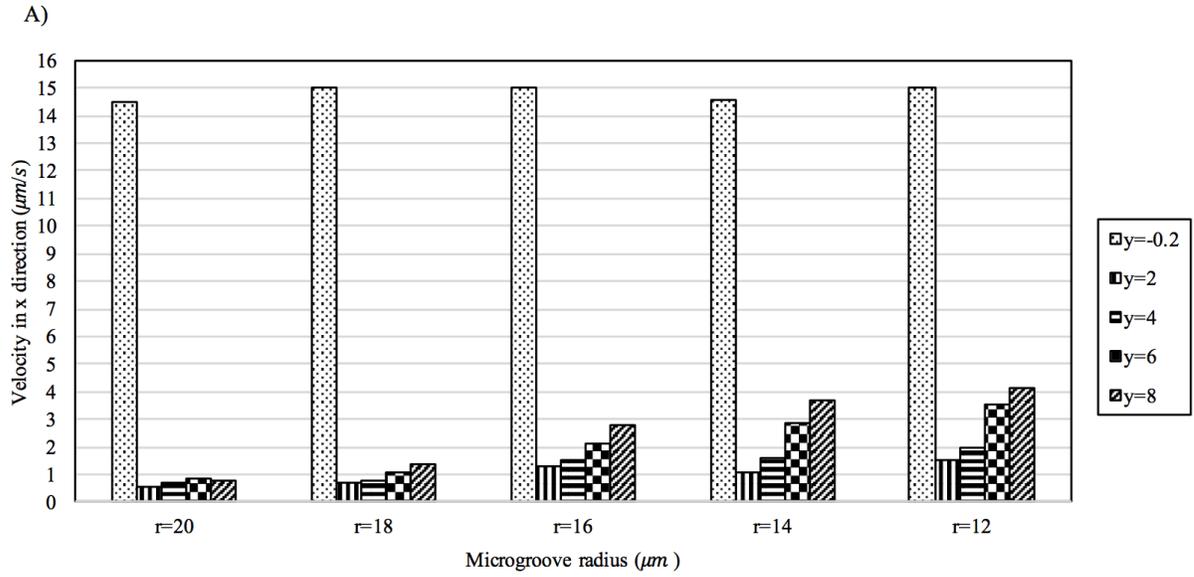

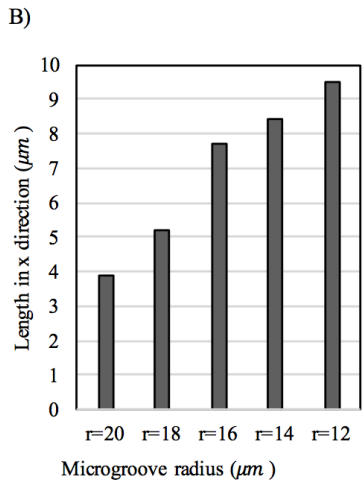

Fig. 7. A) Velocity of fluid at five points in the assumed streamline in semicircular microgrooves in x direction; B) Length of the last point in the assumed streamline in semicircular microgrooves in x direction.

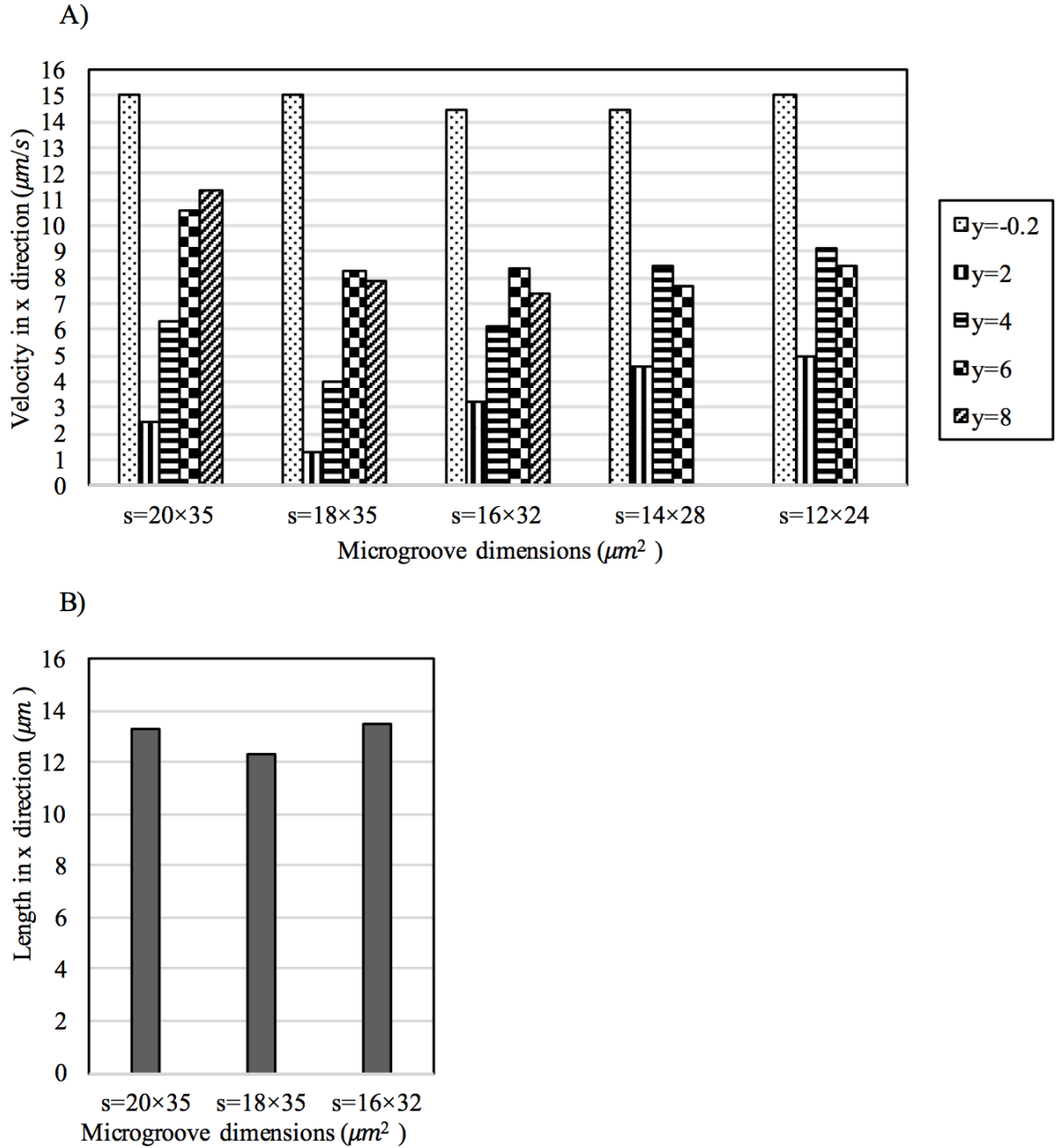

Fig. 8. A) Velocity of fluid at the points in the assumed streamline in rectangular microgrooves in x direction; B) Length of the last point in the assumed streamline in rectangular microgrooves in x direction.

In addition, when the ratio of the length of the last assumed point in x direction to length of microgroove is low, the cell can traverse a longer distance in x direction with a low velocity inside the microgroove, and the cell can have longer time to move toward the bottom of the microgroove. In this way, the probability of cell attachment can increase. As observed in in Fig. 7A and Fig.8A, in the x direction, the velocities of assumed streamline inside the rectangular microgrooves are considerably larger than those of the corresponding semicircular microgrooves due to the sharp corners of the rectangle, but velocities in y direction do not have considerable differences. Further, in the rectangular microgrooves, the ratio of length of the last point in x direction to the length of microgroove is larger than this ratio in semicircular microgrooves. With elevation of the radius of semicircular microgrooves, the velocity in x direction decreases. Also, the ratio of length of the last point in x direction to the diameter of microgroove diminishes. Therefore, it is more possible that the cell enter the recirculation area due to the lower drag force in x direction of the larger semicircular microgroove.

We also evaluated the rectangular microgrooves with recirculation area on the corners of the microgrooves. The results revealed that the velocity in the assumed streamline was larger than the velocity in the assumed rectangular microgrooves in Fig. 8A. Due to the very low recirculation area in the semicircular microgroove with the radius larger than 20 μm, the possibility of initial cell adhesion inside these microgrooves can be very low. In addition, the velocity in x direction inside these microgroove was positive, and the moving cell experienced drag force in x position, which can worsen the cell attachment to the surface.

### 3.3 *Fluid and adhered cell interaction*

After cell adhesion to the receptors coated on the surface, the fluid flow force can detach the cell from the surface, so the interaction of the cell and flowing fluid is important. Therefore, fluid shear stress on the cell plays a pivotal role in cell detachment, and the direction of the flow affects the direction of the shear stress on the cell surface. The recirculating flow has a negative shear stress on the cell, so the percentage of cell surface which experiences negative shear stress can be important in the cell deformation and detachment. In addition, average shear stress on the cell surface can be a representative for evaluating the cell detachment from the surface. Fig. 9 represents the numerical simulation of adhered single cell and flowing fluid interaction. This figure shows the fluid velocity field, recirculating fluid flow near the adhered cell, and shear stress on the cell surface. Fig. 9A and Fig. 9B illustrate the interaction of fluid flow and adhered cells in the rectangular and semicircular microgrooves which have the capacity of single cell docking in x direction. We also assessed the distances between the microgrooves; when the gap between microgrooves was lower than 40 μm, the fluid velocity fluctuated due to the interaction with the cell inside the microgroove affecting the shear stress on the cell inside the next microgroove. This also depends on the cell size, and if the assumed cell size become smaller, the distance between microgrooves can decrease.

We also studied the fluid interaction with attached cell in the flat microchannel. Previous experimental studies have discovered the adhesion force of different types of cells such as different leukocytes ligands and special proteins (receptors) in-vitro. Different cellular ligands and protein bonds were broken when the micropipette aspiration force was between 45 pN and 80 pN [26, 27]. In the flat microchannel as well as the gap between microgrooves, cells surface experienced an average shear stress of about 96 mPa which is shown in Fig. 10B. The resultant shear force in the center of the cell was 120.6 pN, which is larger than the detachment aspiration force. So, in this resultant shear force related to the inlet velocity, the cells cannot adhere to the flat surface, and the gaps between microgrooves can be empty of adhered cells. Indeed, when the inlet velocity is between $10^{-4}$ m/s and $7\times10^{-4}$ m/s, the cells cannot attach to the gap between microgrooves. Also, the resultant shear force on the cells inside microgrooves is lower than micropipette aspiration force causing detachment. For smaller inlet velocities, the cells can attach to the gaps between microgrooves, where the fluid velocity fluctuates due to the interaction with attached cells influencing the shear stress on the cell inside the next microgroove. Further, as the velocity declines, the experiment time is lengthened.

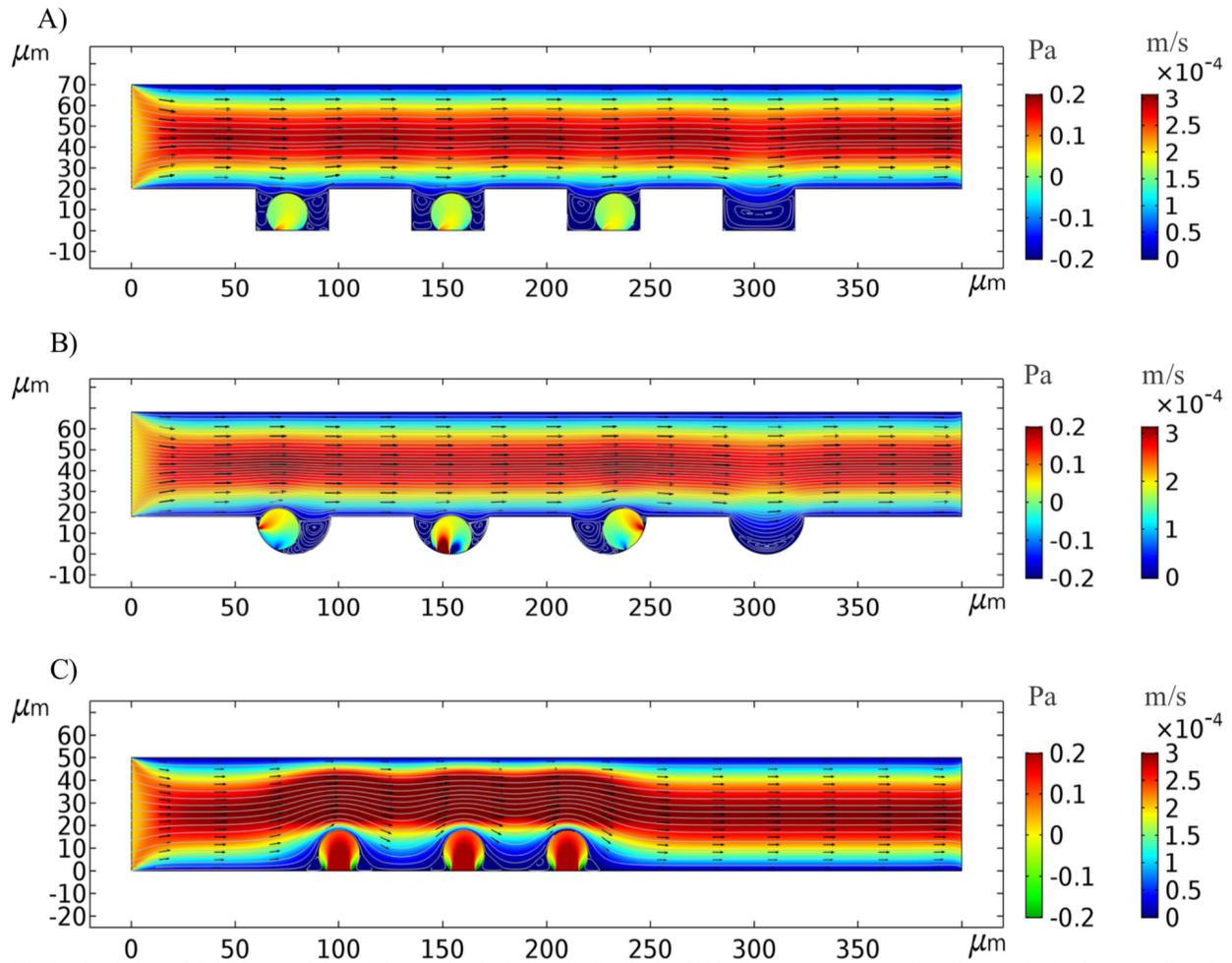

Fig. 9. Simulation of fluid flow and the cell interaction including fluid velocity field, shear stress on the cell surface, and velocity streamlines in: A) Rectangular microgrooves with heights of 24 µm and length of 45 µm; B) Semicircular microgrooves with radii of 18 µm; C) Flat microchannel. The left color bar represents the shear stress between -0.2 Pa and 0.2 Pa, and the right color bar shows the velocity in x direction.

Fig. 10, Fig. 11 and Fig. 12 display the percentage of negative shear stress and average shear stress on the docked cell surface inside the rectangular and semicircular microgrooves, respectively. Due to the large differences of shear stress on the cell across different positions inside the semicircular microgroove, we assessed the cell and fluid interaction in three positions (left, right, and middle) for the cell docking.

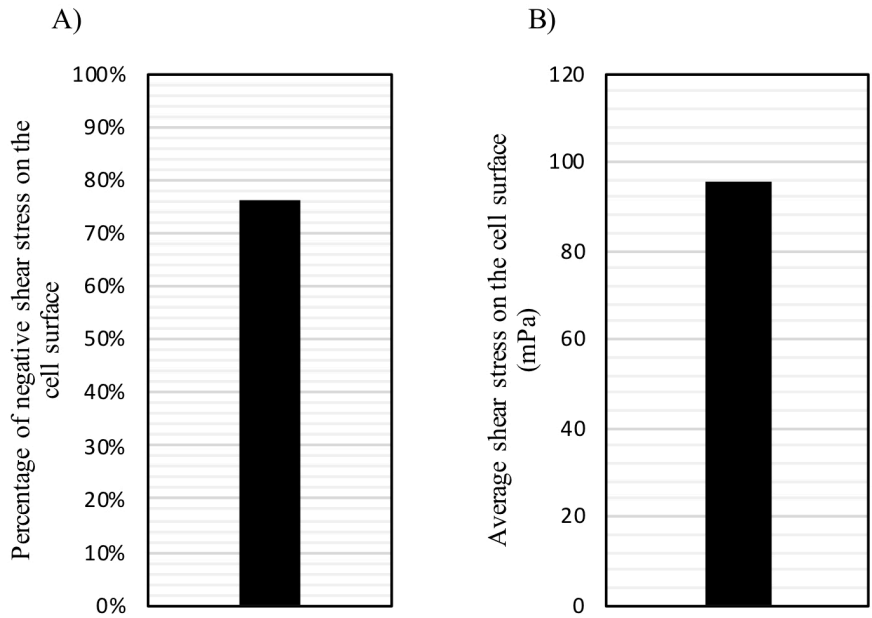

Fig. 10. In the flat microchannel: A) Percentage of negative shear stress on the cell surface; B) Average shear stress on the cell surface.

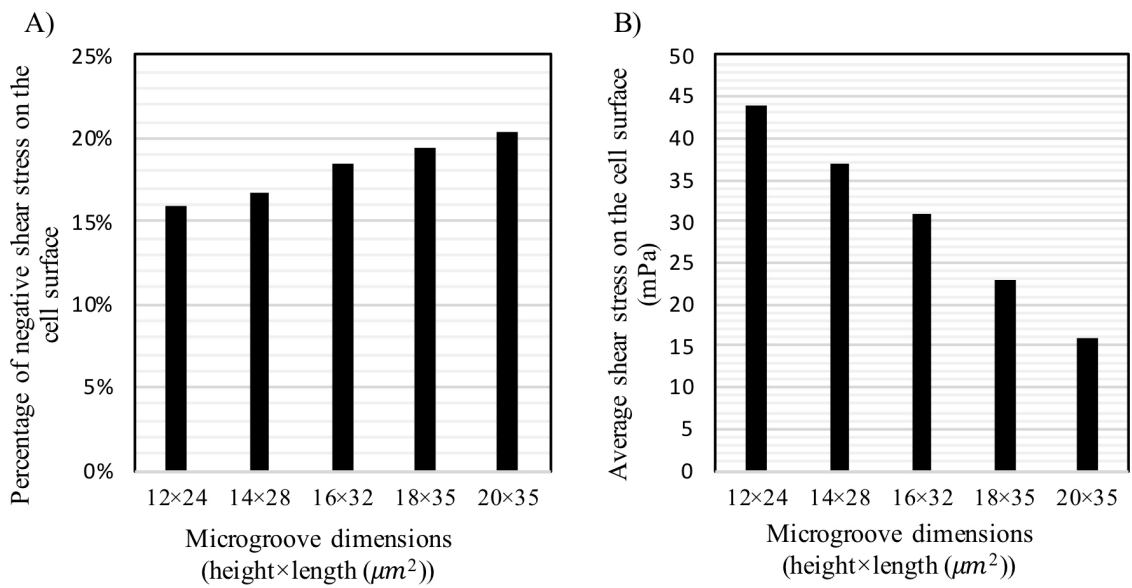

Fig. 11. In rectangular microgrooves: A) Percentage of negative shear stress on the cell surface; B) Average shear stress on the cell surface.

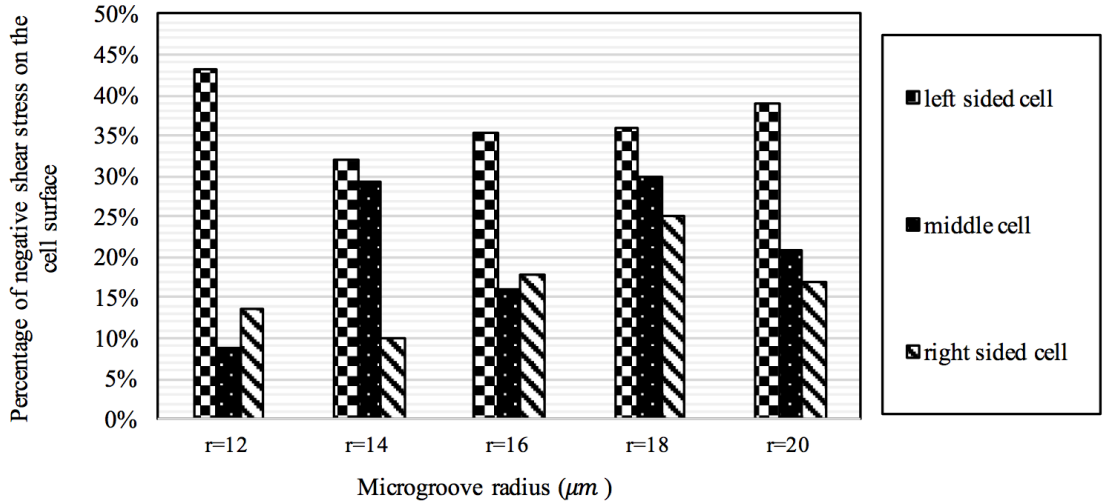

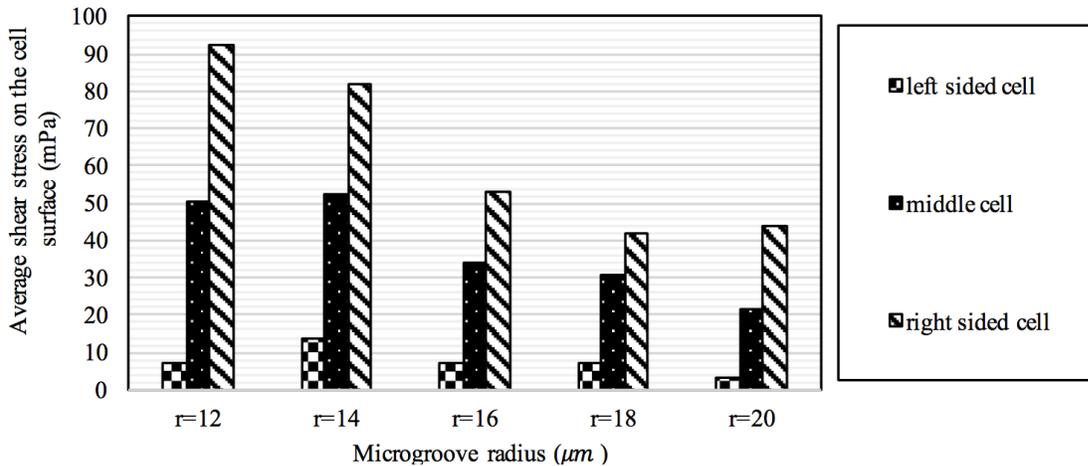

Fig. 12. In semicircular microgrooves: A) Percentage of negative shear stress on the cell surface in three possible position of the cell; B) Average shear stress on the cell surface in three possible position of the cell.

Lower average shear stress and negative shear stress percentage near 50% on the cell surface are in favor of cell attachment. According to the shear stress distribution on the cell surface, the negative and positive shear stresses have occurred on the right and left sides of the docked cell. So, this stress distribution can cause minor cell deformation in the direction of fluid flow. In semicircular microgrooves, the possibility of the cell attachment is high due to the cell docking on the left and right sides of the microgroove. In different rectangular microgrooves, as shown in Fig. 11, the percentage of negative shear stress on the cell surface does not fluctuate. On the other hand, by increasing the length and height of rectangular microgrooves, the average shear stress on the cell surface diminishes. When the cell docks in the left side of the semicircular microgroove, the cell surface experiences a lower average shear stress and higher percentage of negative shear stress. On the other hand, the right sided cell detachment is highly probable due to higher average shear stress (Fig. 12B).

In order to validate these simulation results, there was one similar publication which has investigated experimentally the spheroid formation in microwell-based microfluidics[33]. They have evaluated round bottom wells and flat bottom wells for uniform cell seeding and uniformly sized spheroids. Spheroid formation and growth were dependent on cells adhesion to each other and receiving fresh culture medium, respectively. They have illustrated that hemispherical microstructure are appropriate for guiding uniformly size spheroid formation. When a number of cells get trapped within a hemispherical geometry, the resultant microforce exerted on the cells by the round wall can improve spheroid compaction since the force focuses on the center. In this study, the primary goal of design in these devices is to avoid the spheroids from high shear stress to remove any chance of cell death caused by it. Similarly, we assessed the fluid shear stress effect on the attached cell that can cause cell detachment.

## 4 Conclusion

The ratio of cell radius to the microgroove length and height is important in the possibility of cell docking. In this study, microgroove sizes were assumed according to single cell docking along the length of microchannel in x direction with a radius of 10 μm. It was concluded that three parameters play a pivotal role in the cell capturing by the coated receptors on the surface. Firstly, maximum and minimum height of recirculation area should be high enough, though the height of recirculation area should not exceed the adhered cell height (18 μm). Higher recirculation areas can enhance the possibility of cells detachment from fluid flow and cells attachment to the surface receptors in microgrooves, but higher recirculation area can whelm a larger volume of cells and reduce the delivery of nutrients to the cells as well as cellular waste disposal. So, the recirculation areas' heights should not be higher than 18 μm. Secondly, the drag force as well as fluid velocity in the flow direction (x direction) exerted on the moving cell inside the microgroove should be low. Finally, the cell should move slowly with low velocity in x direction with a long distance inside the microgroove. If the microgroove geometry has these three parameters, the cell attachment can be highly possible. Although in the large rectangular microgroove (s=20×35) the recirculation area was high enough, the drag force in x direction was high due to the large velocity in x direction. On the other hand, in the large semicircular microgroove (r=18 or 20 μm) in favor of cell docking, the recirculation area was large, and the drag force in x direction was low.

After initial adhesion of the cell, there can be two parameters against cell detachment. First, lower average shear stress on the cell surface is in favor of cell docking inside the microgroove. Secondly, the percentage of negative shear stress close to 50% as well as negative and positive stresses distribution on the left and right side of the cell surface are favorable for cell attachment. In semicircular microgrooves, the negative stress distribution was on the right side of the cell due to the fluid recirculation area. On the contrary, in rectangular microgrooves, the negative shear stress distribution was on the left and right sides at the bottom of the cell; therefore, the top of the cell experiences positive shear stress in flow direction leading to a high probability of cell deformation. When the cell is captured in the recirculation area, the flow direction changes; hence, the possibility of cell docking on the left side of the microgrooves may be higher than on the other sides. In semicircular microgrooves, the cells docked on the left side experience lower average shear stress and have a large percentage of negative shear stress. Consequently, the probability of cell detachment can be low.

The authors declare that they have no conflict of interest.

**References**

1. Gossett, D.R., et al., *Label-free cell separation and sorting in microfluidic systems.* Analytical and bioanalytical chemistry, 2010. **397**(8): p. 3249-3267.
2. Di Carlo, D., L.Y. Wu, and L.P. Lee, *Dynamic single cell culture array.* Lab on a Chip, 2006. **6**(11): p. 1445-1449.
3. Polacheck, W.J., et al., *Microfluidic platforms for mechanobiology.* Lab on a Chip, 2013. **13**(12): p. 2252-2267.
4. Zhou, J. and L.E. Niklason, *Microfluidic artificial "vessels" for dynamic mechanical stimulation of mesenchymal stem cells.* Integrative Biology, 2012. **4**(12): p. 1487-1497.
5. Vining, K.H. and D.J. Mooney, *Mechanical forces direct stem cell behaviour in development and regeneration.* Nature Reviews Molecular Cell Biology, 2017. **18**(12): p. 728.
6. Khademhosseini, A., et al., *Co-culture of human embryonic stem cells with murine embryonic fibroblasts on microwell-patterned substrates.* Biomaterials, 2006. **27**(36): p. 5968-5977.
7. Sackmann, E.K., A.L. Fulton, and D.J. Beebe, *The present and future role of microfluidics in biomedical research.* Nature, 2014. **507**(7491): p. 181.
8. Khademhosseini, A., et al., *Cell docking inside microwells within reversibly sealed microfluidic channels for fabricating multiphenotype cell arrays.* Lab on a Chip, 2005. **5**(12): p. 1380-1386.
9. Kumar, A. and A. Bhardwaj, *Methods in cell separation for biomedical application: cryogels as a new tool.* Biomedical materials, 2008. **3**(3): p. 034008.

10. Weibel, D.B. and G.M. Whitesides, *Applications of microfluidics in chemical biology.* Current opinion in chemical biology, 2006. **10**(6): p. 584-591.
11. Tian, W.-C. and E. Finehout, *Microfluidics for biological applications*. Vol. 16. 2009: Springer Science & Business Media.
12. Koh, W.-G., A. Revzin, and M.V. Pishko, *Poly (ethylene glycol) hydrogel microstructures encapsulating living cells.* Langmuir, 2002. **18**(7): p. 2459-2462.
13. Khademhosseini, A., et al., *A soft lithographic approach to fabricate patterned microfluidic channels.* Analytical chemistry, 2004. **76**(13): p. 3675-3681.
14. Fidkowski, C., et al., *Endothelialized microvasculature based on a biodegradable elastomer.* Tissue engineering, 2005. **11**(1-2): p. 302-309.
15. Vahey, M.D. and J. Voldman, *An equilibrium method for continuous-flow cell sorting using dielectrophoresis.* Analytical chemistry, 2008. **80**(9): p. 3135-3143.
16. Petersson, F., et al., *Free flow acoustophoresis: microfluidic-based mode of particle and cell separation.* Analytical chemistry, 2007. **79**(14): p. 5117-5123.
17. Moehlenbrock, M.J., A.K. Price, and R.S. Martin, *Use of microchip-based hydrodynamic focusing to measure the deformation-induced release of ATP from erythrocytes.* Analyst, 2006. **131**(8): p. 930-937.
18. Schaff, U.Y., et al., *Vascular mimetics based on microfluidics for imaging the leukocyte–endothelial inflammatory response.* Lab on a Chip, 2007. **7**(4): p. 448-456.
19. Manbachi, A., et al., *Microcirculation within grooved substrates regulates cell positioning and cell docking inside microfluidic channels.* Lab on a Chip, 2008. **8**(5): p. 747-754.
20. Cioffi, M., et al., *A computational and experimental study inside microfluidic systems: the role of shear stress and flow recirculation in cell docking.* Biomedical microdevices, 2010. **12**(4): p. 619-626.
21. Khabiry, M., et al., *Cell docking in double grooves in a microfluidic channel.* Small, 2009. **5**(10): p. 1186-1194.
22. Khabiry, M. and N. Jalili, *A Microfluidic Platform Containing Sidewall Microgrooves for Cell Positioning and Trapping.* Nanobiomedicine, 2015. **2**: p. 4.
23. Park, J., et al., *Microfabricated grooved substrates as platforms for bioartificial liver reactors.* Biotechnology and bioengineering, 2005. **90**(5): p. 632-644.
24. Yamada, M. and M. Seki, *Hydrodynamic filtration for on-chip particle concentration and classification utilizing microfluidics.* Lab on a Chip, 2005. **5**(11): p. 1233-1239.
25. Duprat, C. and H.A. Shore, *Fluid-Structure Interactions in Low-Reynolds-Number Flows*. 2015: Royal Society of Chemistry.
26. Wang, Y.-l. and D.E. Discher, *Cell mechanics*. Vol. 83. 2007: Academic Press.
27. Ethier, C.R. and C.A. Simmons, *Introductory biomechanics: from cells to organisms*. 2007: Cambridge University Press.
28. Fung, Y.-c., *Biomechanics: circulation*. 2013: Springer Science & Business Media.
29. Gaver, D.P. and S.M. Kute, *A theoretical model study of the influence of fluid stresses on a cell adhering to a microchannel wall.* Biophysical journal, 1998. **75**(2): p. 721-733.
30. Païdoussis, M.P., S.J. Price, and E. De Langre, *Fluid-structure interactions: cross-flow-induced instabilities*. 2010: Cambridge University Press.
31. Reddy, J.N., *An introduction to continuum mechanics*. 2013: Cambridge university press.
32. Fung, Y.-c., *Biomechanics: mechanical properties of living tissues*. 2013: Springer Science & Business Media.
33. Moshksayan, K., et al., *Spheroids-on-a-chip: Recent advances and design considerations in microfluidic platforms for spheroid formation and culture.* Sensors and Actuators B: Chemical, 2018. **263**: p. 151-176.